\documentclass[%
 reprint,
 superscriptaddress,showkeys,
 amsmath,amssymb,
 aps,
]{revtex4-1}

\usepackage{graphicx}
\usepackage{dcolumn}
\usepackage{bm}
\usepackage{xcolor}
\usepackage{url}
\usepackage{makecell}
\usepackage{subfigure}
\usepackage{multirow}
\usepackage{booktabs}
\begin{document}


\title{Exclusive production of $T^{-}_{cc}$ in hadron-hadron ultraperipheral collisions}

\author{Xiao-Peng Wang}
\affiliation{Institute of Modern Physics, Chinese Academy of Sciences, Lanzhou 730000, China}
\affiliation{School of Nuclear Science and Technology, Lanzhou University, Lanzhou 730000, China}
\affiliation{School of Nuclear Science and Technology, University of Chinese Academy of Sciences, Beijing 100049, China}



\author{Ya-Ping Xie}
\affiliation{Institute of Modern Physics, Chinese Academy of Sciences, Lanzhou 730000, China}
\affiliation{School of Nuclear Science and Technology, University of Chinese Academy of Sciences, Beijing 100049, China}

\author{Yin Huang}
\email{ huangy2019@swjtu.edu.cn (Corresponding author)}
\affiliation{School of Physical Science and Technology, Southwest Jiaotong University, Chengdu 610031, China}

\author{Xu-Rong Chen}
\email{xchen@impcas.ac.cn (Corresponding author)}
\affiliation{Institute of Modern Physics, Chinese Academy of Sciences, Lanzhou 730000, China}
\affiliation{School of Nuclear Science and Technology, University of Chinese Academy of Sciences, Beijing 100049, China}

\date{\today}

\begin{abstract}
Understanding the production mechanisms and utilizing them as probes to investigate the structure of exotic states represent some of
the most actively studied research areas in particle physics.  In this study, we present a theoretical analysis of the charmed meson
$T^{-}_{cc}$ in the $\gamma{}p\to{}D^{+}T^{-}_{cc}\Lambda^{+}_{c}$ and $\gamma\gamma\to{}D^{+}T^{-}_{cc}D^{*0}$ reactions, considering
$T^{+}_{cc}$ as a $DD^{*}$ molecule.  The differential cross-section and total cross-section for the photoproduction of $T^{-}_{cc}$ in
the $\gamma{}p\to{}D^+T^{-}_{cc}\Lambda^{+}_{c}$ and $\gamma\gamma\to{}D^{+}T^{-}_{cc}D^{*0}$ reactions are presented for nucleus-nucleus
(nucleon-nucleon) ultraperipheral collisions at HL-LHC and RHIC, respectively.  Taking into account the integrated luminosity per typical run
and the luminosity of photons from the nucleus (nucleon), we observe a significant event count for $T^{-}_{cc}$ production in p-p ultraperipheral
collisions at HL-LHC.

\end{abstract}

\maketitle

\section{Introduction}\label{sec:intro}
The quark model offers a convenient framework for classifying hadrons, effectively encompassing the majority of hadronic states.
Nonetheless, significant experimental advancements have been made in recent years, leading to the observation of numerous exotic hadrons~\cite{Guo:2017jvc,Chen:2022asf}.
These exotic mesons exhibit an internal structure more complex than the simple  $q\bar{q}$ configuration for mesons or $qqq$ configuration
for baryons in the traditional constituent quark models.

The study of exotic hadrons has a long and storied history. However, it entered a new and exciting era in 2003 when the Belle collaboration
made a groundbreaking discovery of the $X(3872)$ in the $\pi^{+}\pi^{-}J/\psi$ mass spectra~\cite{Belle:2003nnu}. The $X(3872)$, based on its
observed decay mode, is known to consist of at least four distinct valence quarks, making it a candidate for an exotic hadron.
Another well-known exotic hadron candidate is the charged-particle $Z_c^+(3900)$, which was initially observed by the BESIII collaboration
in the $\pi^{\pm}J/\psi$ mass spectrum~\cite{BESIII:2013ris}. This observation was later confirmed by the Belle collaboration in the process $e^{+}e^{-}\to\pi^{+}\pi^{-}J/\psi$~\cite{Belle:2013yex}. Moreover, the LHCb Collaboration has made significant strides in the field by
reporting several hidden-charm pentaquark states \cite{LHCb:2015yax,LHCb:2020jpq}.

These discoveries have garnered significant attention towards exotic hadron states, particularly those containing charm quarks. Notably, it brings to mind the double-charm meson $T^{+}_{cc}$, which was observed by the LHCb Collaboration in the $D^0D^0\pi^{+}$ invariant mass spectrum~\cite{LHCb:2021auc}.  The $T^{+}_{cc}$ meson's mass, width, and quantum numbers were precisely measured as follows:
\begin{align}
M&=3875.09~~{\rm MeV}+\delta{}m,\nonumber\\
\Gamma&=48\pm2^{+0}_{-14}~~{\rm KeV},~~I(J^{P})=0(1^{+}).
\end{align}
Since the mass of $T^+_{cc}$ lies just below the nominal $D^{*+}D^0$ threshold, it can be interpreted as a hadronic molecule \cite{Albaladejo:2021vln, Montesinos:2023qbx,Du:2021zzh,Agaev:2021vur,PhysRevD.104.L051502,Yan:2021wdl,Ling:2021bir,Wang:2023ovj}.  It was worth noting that the compact multi-quark structure for
$T^+_{cc}$ are also proposed in Refs.~\cite{Kim:2022mpa,Eichten:2017ffp}.

Theoretical investigations on production mechanisms and further experimental information on production cross section will be helpful to distinguish which inner structure
of the $T^+_{cc}$ state is possible.  This dependence arises primarily from the fact that the different yields are strongly influenced by the internal structure of the
hadrons.  Presently, the photoproduction of $T^{-}_{cc}$, which serves as the antiparticle to $T^{+}_{cc}$, has undergone investigation as documented in Ref.~\cite{PhysRevD.104.116008}.  The process $\gamma p\rightarrow D^+T^{-}_{cc}\Lambda^+_c$ involves the utilization of the central diffractive mechanism.  In their
consideration, the reaction channel involves the exchange of $D^{(*)}$ mesons in the $t$-channel, while the $s$- and $u$-channels are significantly suppressed due to the
involvement of two additional $c\bar{c}$ pair creation.  Their findings reveal that the total cross-section for $\gamma p\rightarrow D^+T^{-}_{cc}\Lambda^+_c$ is
approximately 1 $\rm Pb$.

When compared to the lower production cross-section of the $\gamma p\rightarrow D^+T^{-}_{cc}\Lambda^+_c$ reaction shown in Ref.~\cite{PhysRevD.104.116008},
ultraperipheral collisions (UPCs) can significantly increase the probability of $T^-_{cc}$ production~\cite{Esposito:2021ptx,Bertulani:2005ru,Baur:2001jj,Krauss:1997vr,Baltz:2007kq}.
In UPCs, electromagnetic interactions dominate, occurring when the impact parameter of two ions exceeds the sum of their radii.  By employing the $\rm Weizs\ddot{a}cker-Williams$ method \cite{PhysRev.45.729,Bertulani:2005ru,Baltz:2007kq}, the electromagnetic field originating from highly-charged nuclei can be treated as an equivalent flux of photons.
As the photon flux is directly proportional to the charge number of ions, highly-charged ions offer a substantial photon number density.
This suggests that if the $\gamma p\rightarrow D^+T^{-}_{cc}\Lambda^+_{c}$ reaction occurs in ultraperipheral collisions, there will be a significant increase in
the probability of $T^-_{cc}$ production.  Moreover, we also proposed to observe two-photon scattering \cite{Baltz:2007kq} ($\gamma \gamma \rightarrow D^+ T^-_{cc}D^*$) as another part of the search for the $T^-_{cc}$ in UPCs due to the fact that  $Z(3930)$, $X(3915)$, and $X(4350)$ were observed in this process by Belle collaboration \cite{Belle:2005rte,Belle:2009and,Belle:2009rkh}.  As a result, UPCs serve as a crucial platform for investigating the photoproduction of  the $T^-_{cc}$~\cite{Klein:2019avl,Xie:2021sik}.

This paper is organized as follows.  Theoretical frameworks for the production of $T^{-}_{cc}$ in $\gamma p\rightarrow D^+T^{-}_{cc}\Lambda^+_c$ and $\gamma \gamma \rightarrow D^+ T^-_{cc}D^*$ in UPCs are presented in Section$\,$\ref{sec:theo}, respectively. The numerical calculations are given in Section$\,$\ref{sec:numerical redult}.  Finally, a conclusion is given in section$\,$\ref{sec:summary}.

\section{THEORETICAL FRAMEWORK }\label{sec:theo}
\begin{figure*}[hbtp]
	\centering
	\subfigure[\label{fig:subfig:single1}]{
		\includegraphics[scale=0.3]{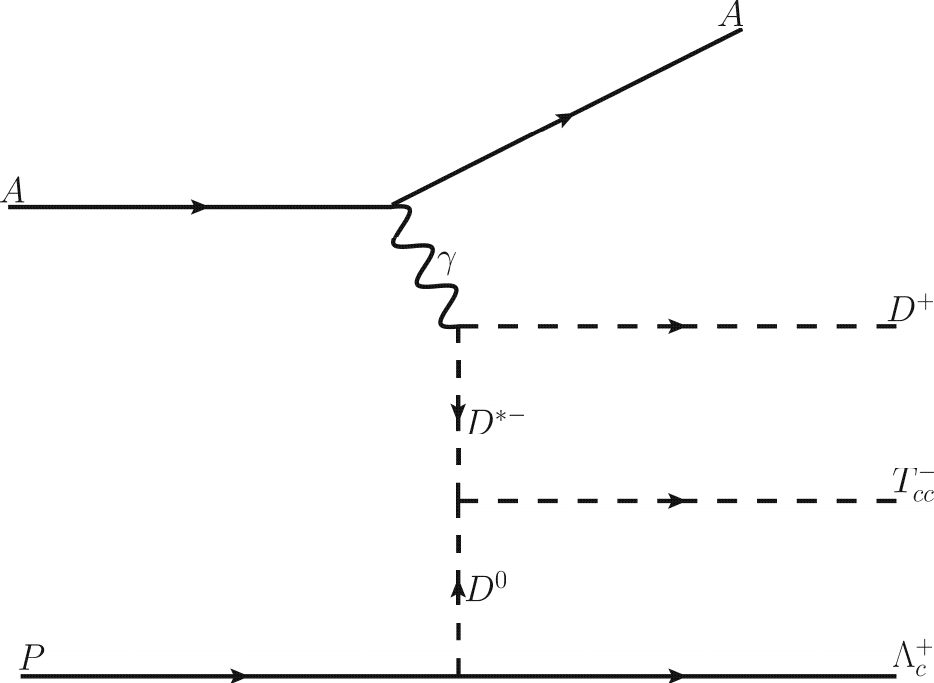}}
	\hspace{0.0098in}
	\subfigure[\label{fig:subfig:single2}]{
		\includegraphics[scale=0.3]{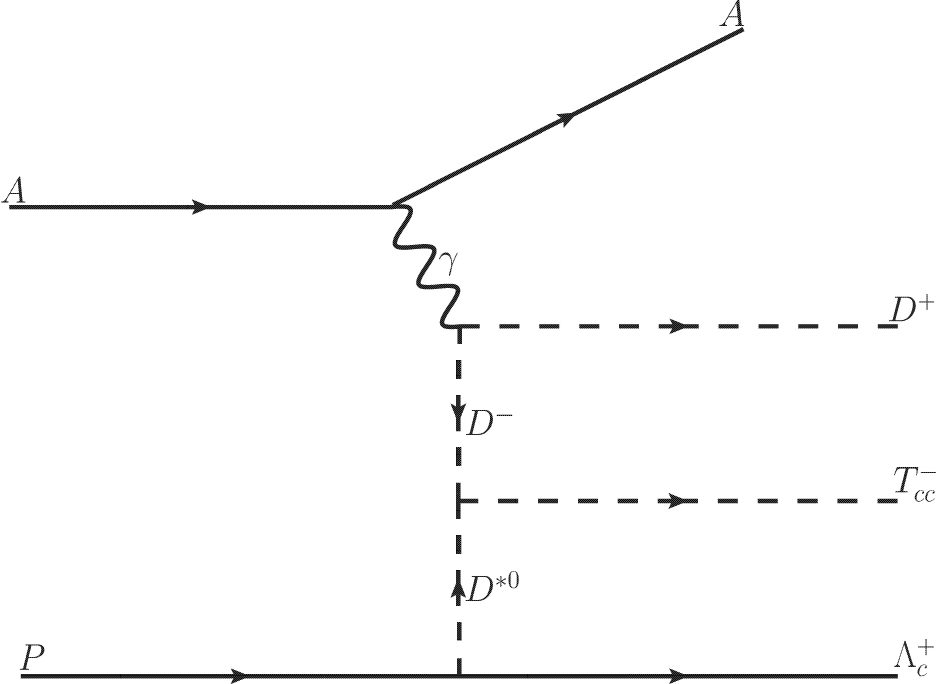}}
	\centering
	\hspace{0.0098in}
	\subfigure[\label{fig:subfig:single3}]{
		\includegraphics[scale=0.3]{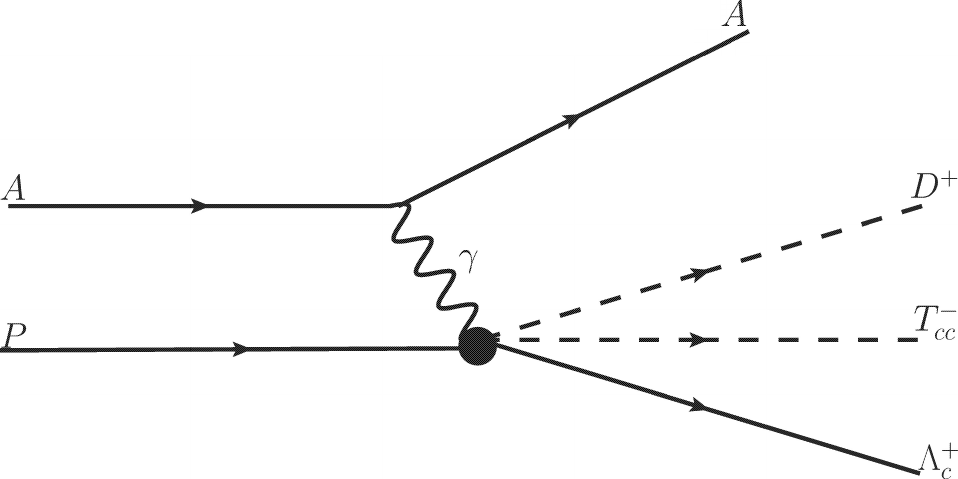}}
	
	\caption{Production of $p +A\rightarrow A+ D^{+}+T^{-}_{cc}+\Lambda^+_c$ and $p +p\rightarrow p+ D^{+}+T^{-}_{cc}+\Lambda^+_c$ in pA or pp UPCs.}
	\label{fig:single_gama}
\end{figure*}

In this work, we investigate the production of $T^{-}_{cc}$ through one-photon and two-photon processes of ultra-peripheral collisions (UPCs). The corresponding Feynman
diagrams are depicted in Figs.~\ref{fig:single_gama} and \ref{fig:double_gama}. We can find that the high-luminosity photon flux is first emitted from the nucleus or nucleon, resulting in the creation of a pair of high-energy $\bar{D}^{*}\bar{D}$ mesons.  Due to the attractive interaction between $\bar{D}^{*}$ and $\bar{D}$, a $T^{-}_{cc}$ molecule
is formed in the final state.  Next, we will discuss in detail the production mechanisms of $T^{-}_{cc}$, corresponding to Figs.~\ref{fig:single_gama} and \ref{fig:double_gama}, respectively.

\subsection{The production of $T^-_{cc}$ in one-photon process}
 Within the framework of UPCs, the production cross-section of $A(p) p\rightarrow D^+  T^-_{cc}\Lambda^+_{c}A(p)$ reaction was given
\cite{KLEIN2017258,Xie:2020wfe},
\begin{equation}
	\sigma(A p\rightarrow A D^+T^-_{cc}\Lambda_c^+ )=\int \mathrm{d} k \frac{\mathrm{d} N_\gamma(k)}{\mathrm{d} k} \sigma_{\gamma p \rightarrow D^+  T^-_{cc}\Lambda^+_{c}}(\rm W),
	\label{eq:single_gama_totsigma}
\end{equation}
where $\frac{\mathrm{d} N_\gamma(k)}{\mathrm{d} k}$ is photon flux, with the $k$ representing the energy of the photon emitted from the nucleus (nucleon).  $\rm W$ is
the center of mass energy of the photon and proton system.  Note that ${\rm W}=\sqrt{2k\sqrt{s}}$ can be determined based on $k$ and the total energy $s$ of the system.
Simplifying, we obtain the $\rm W$-dependent differential cross-section
\begin{equation}
	\frac{\mathrm{d}\sigma}{\mathrm{d}{\rm W}}=\left(\frac{\mathrm{d} N_\gamma(k)}{\mathrm{d} k}\frac{{\rm W}}{\sqrt{s}}\right) \sigma_{\gamma p \rightarrow   D^+T^-_{cc}\Lambda_c^+}({\rm W}).
	\label{eq:single__gama_diff_sigma}
\end{equation}

In order to make a reliable prediction for the cross-section of the $A(p) p\rightarrow D^+ T^-_{cc}\Lambda^+_c A(p)$ reaction,
we need to address the follow two key issues: the value of the photon flux and the $\sigma_{\gamma p \rightarrow D^+T^-_{cc}\Lambda_c^+}({\rm W})$.
The photon flux from the nucleus is described by the equation~\cite{PhysRevC.60.014903}
\begin{small}
	\begin{equation}
		\frac{\mathrm{d} N_\gamma(k)}{\mathrm{d} k}\!=\!\frac{2 Z^2 \alpha_{em}}{\pi k}\!\left(\! \xi K_0(\xi) K_1(\xi)\!-\!\frac{\xi^2}{2}\left[K_1^2(\xi)\!-\! K_0^2(\xi)\right]\!\right)\!,\!
		\label{eq:single_gamma_nuclus_photon_flux}
	\end{equation}
\end{small}
where $K_0$ and $K_1$ are modified Bessel functions, $Z$ is the ion charge, $\alpha_{em}=1 / 137$, $\xi=b_{\min } k / \gamma_{\mathrm{L}}$ with
$\gamma_{\mathrm{L}}=\sqrt{s}/(2m_p)$ represents the Lorentz boost factor. The value of $b_{\min}=R_A+R_p$ is the sum of the nucleus and proton
charge radius, where $R_A$ is often defined as~\cite{Lappi:2010dd}
\begin{equation}
	R_{A}= (1.12A^{1/3}-0.86A^{-1/3})\,(\rm fm).
\end{equation}
It is important to note that the photon flux emitted from the proton differs from the photon flux from the nucleus, and it can be expressed using
the dipole form factor~\cite{Drees:1988pp,Bertulani:2005ru}
\begin{equation}
	\begin{split}
		\frac{d n}{d k}(k)=&\frac{\alpha_{e m}}{2 \pi k}\left[1+\left(1-\frac{2 k}{\sqrt{s}}\right)^{2}\right]\times\\
		&\left(\ln \Omega-\frac{11}{6}+\frac{3}{\Omega}-\frac{3}{2 \Omega^{2}}+\frac{1}{3 \Omega^{3}}\right),
	\end{split}
	\label{eq:proton_gama_flux}
\end{equation}
where $\Omega=1+0.71 ,\mathrm{GeV^2}/Q^2_{\mathrm{min}}$, and $Q^2_{\mathrm{min}}=k^2/\gamma^2_{\mathrm{L}}$ represents the minimum momentum transfer
possible in the reaction.

The differential cross section in the c.m. frame for the $\gamma p \to D^+T^-_{cc}\Lambda_c^{+}$ reaction reads
\begin{align}
&d\sigma(\gamma{}p\to{}D^+T^-_{cc}\Lambda_c^{+})\nonumber\\
&=\frac{1}{(2\pi)^5}\frac{1}{4(k_{1}\cdot{}k_2)}\bar{\sum_{s_i,s_f}}|-i{\cal{M}}(\gamma{}p\to{}D^+T^-_{cc}\Lambda_c^{+})|^2\nonumber\\
&\times{}\frac{d^3\vec{p}_1}{2E_1}\frac{d^3\vec{p}_2}{2E_2}\frac{d^3\vec{p}_3}{2E_3}\delta^4(k_1+k_2-p_1-p_2-p_3)\label{eq7},
\end{align}
where $E_1$, $E_2$,$E_3$ and $p_1,p_2,p_3$ stand for the energies and four momentum of $D^{+}$ ,$\bar{T}^{-}_{cc}$, and $\Lambda_c^{+}$, respectively.
$k_1$ and $k_2$ are the four momentum of the initial photon and proton,respectively, and $m_p$ and $m_{\Lambda^{+}_c}$ are the masses of the proton and
$\Lambda_c^{+}$,respectively.  The ${\cal{M}}(\gamma{}p\to{}D^+T^-_{cc}\Lambda_c^{+})$ represents the total scattering amplitude for the $\gamma p \to D^+T^-_{cc}\Lambda_c^{+}$
reaction, which has be computed in Ref.~\cite{PhysRevD.104.116008}
\begin{align}
-i{\cal{M}}&=\bar{u}(p_3,\lambda_{\Lambda_c^{+}})\sum_{j=a,b,c}{\cal{W}}_j^{\mu\nu}u(k_2,\lambda_p)\nonumber\\
                                                   &\times{}\epsilon_{\nu}(k_1,\lambda_{\gamma})\epsilon^{*}_{\mu}(p_2,\lambda_{\bar{T}^{-}_{cc}}),
\end{align}
with
\begin{align}
{\cal{W}}_a^{\mu\nu}&=-g_{D^{*}D\gamma}g_{DN\Lambda_c}g_{\bar{T}_{cc}}\gamma_5\epsilon_{\alpha\nu\beta\rho}k_1^{\alpha}q_{1}^{\beta}\nonumber\\
                    &\times\frac{-g^{\mu\rho}+q_1^{\mu}q_1^{\rho}/m^2_{D^{*-}}}{q_1^2-m^2_{D^{*-}}}\frac{{\cal{F}}_{\bar{D}^0}{\cal{F}}_{D^{*-}}}{q_2^2-m^2_{\bar{D}^0}}\label{eq9},\\
{\cal{W}}_b^{\mu\nu}&=-ieg_{D^{*}N\Lambda_c}g_{\bar{T}_{cc}}\gamma_{\rho}(q_1^{\nu}-p_1^{\nu})\nonumber\\
                    &\times\frac{-g^{\mu\rho}+q_2^{\mu}q_2^{\rho}/m^2_{\bar{D}^{*0}}}{q_2^2-m^2_{\bar{D}^{*0}}}\frac{{\cal{F}}_{\bar{D}^{*0}}{\cal{F}}_{D^{-}}}{q_1^2-m^2_{D^{-}}}\label{eq10},\\
{\cal{W}}_c^{\mu\nu}&=-i2eg_{D^{*}N\Lambda_c}g_{\bar{T}_{cc}}(-\gamma_{\mu}+\frac{m_p-m_{\Lambda_c^{+}}}{m^2_{\bar{D}^{*0}}}q_{2\mu})\nonumber\\
                    &\times\frac{k_1^{\nu}-p_1^{\nu}}{q_2^2-m^2_{\bar{D}^{*0}}}\frac{{\cal{F}}_{\bar{D}^{*0}}{\cal{F}}_{D^{-}}}{q_1^2-m^2_{D^{-}}}\label{eq11},
\end{align}
where correspond to the Feynman diagrams as well as the contact terms discussed in Ref.~\cite{PhysRevD.104.116008}, respectively.  In the above equation,
$u$ and $\epsilon$ are the Dirac spinor and polarization vector, respectively, and $\lambda$ is the helicities.  Coupling constants $g_{DN\Lambda_c}=-13.98$
and $g_{D^{*}N\Lambda_c}=-5.20$ are computed from the SU(4) invariant Lagrangians in terms of $g_{\pi{}NN}=13.45$ and $g_{\rho}NN=6$ \cite{Dong:2010xv,Huang:2016ygf,Okubo:1975sc}.
$g_{D^{*}D\gamma}=0.173-0.228$ GeV$^{-1}$ is determined by the radiative decay widths of $D^{*}$ \cite{10.1093/ptep/ptaa104}.  The coupling constants
$g_{T_{cc}D^{*+}D^0}=3.67$ GeV and $g_{ T_{cc}D^{0}D^+}=-3.92$ GeV are derived from chiral unitary theory, where $T_{cc}$ is identified as an $S$-wave $D^{*}D$ molecule \cite{Feijoo:2021ppq}.  $e=\sqrt{4\pi\alpha}$ with $\alpha$ being the fine-structure constant,
and $\epsilon^{\mu\nu\alpha\beta}$ is the Levi-Civita tensor.

Considering the internal structure of the exchange mesons, the form factor must be taken into account.  In this work, the monopole form factors ${\cal{F}}_{\bar{D}^{(*)0}}$
and ${\cal{F}}_{D^{(*)-}}$ that can be found in Eqs.~(\ref{eq9}-\ref{eq11}) are utilized, as shown in \cite{PhysRevD.104.116008}
\begin{align}
{\cal{F}}_{i}=\frac{\Lambda_{i}^{2}-m_{i}^{2}}{\Lambda_{i}^{2}-t_{i}}, \quad i=\bar{D}^0, D^{*-}, \label{eq:formfactor}
\end{align}
where $m_i$ and $t_i$ represent the mass and the four-momentum square of exchange mesons $\bar{D}^0$ or $D^{*-}$. The cutoff $\Lambda_{i}=m_i+\alpha\Lambda_{\rm QCD}$,
where $\alpha$ is a parameter related to the nonperturbative property of QCD at the low-energy scale.  In Ref.~\cite{PhysRevD.104.116008}, $\Lambda_{\rm QCD}=0.22,\rm GeV$ is
adopted, and $\alpha=1.5$ or $1.7$ is computed by fitting the experimental data~\cite{Guo:2016iej, BaBar:2006qlj, Belle:2007qxm}.

\subsection{The production of $T^-_{cc}$ in two-photon process}
\begin{figure}[htbp]
	\centering
	\subfigure[\label{fig:subfig:double_gama1}]{
	\includegraphics[scale=0.22]{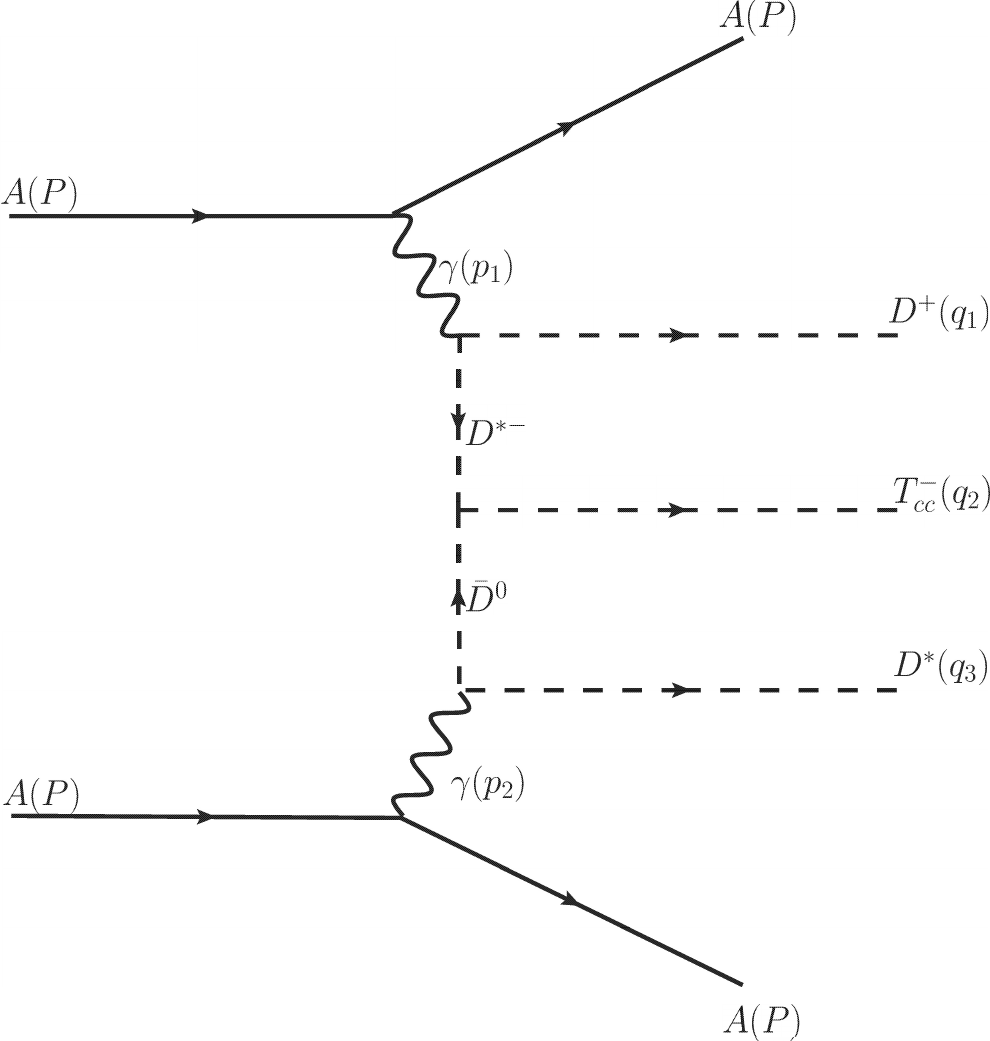}}
	\hspace{0.00098in}
	\subfigure[\label{fig:subfig:double_gama2}]{\includegraphics[scale=0.22]{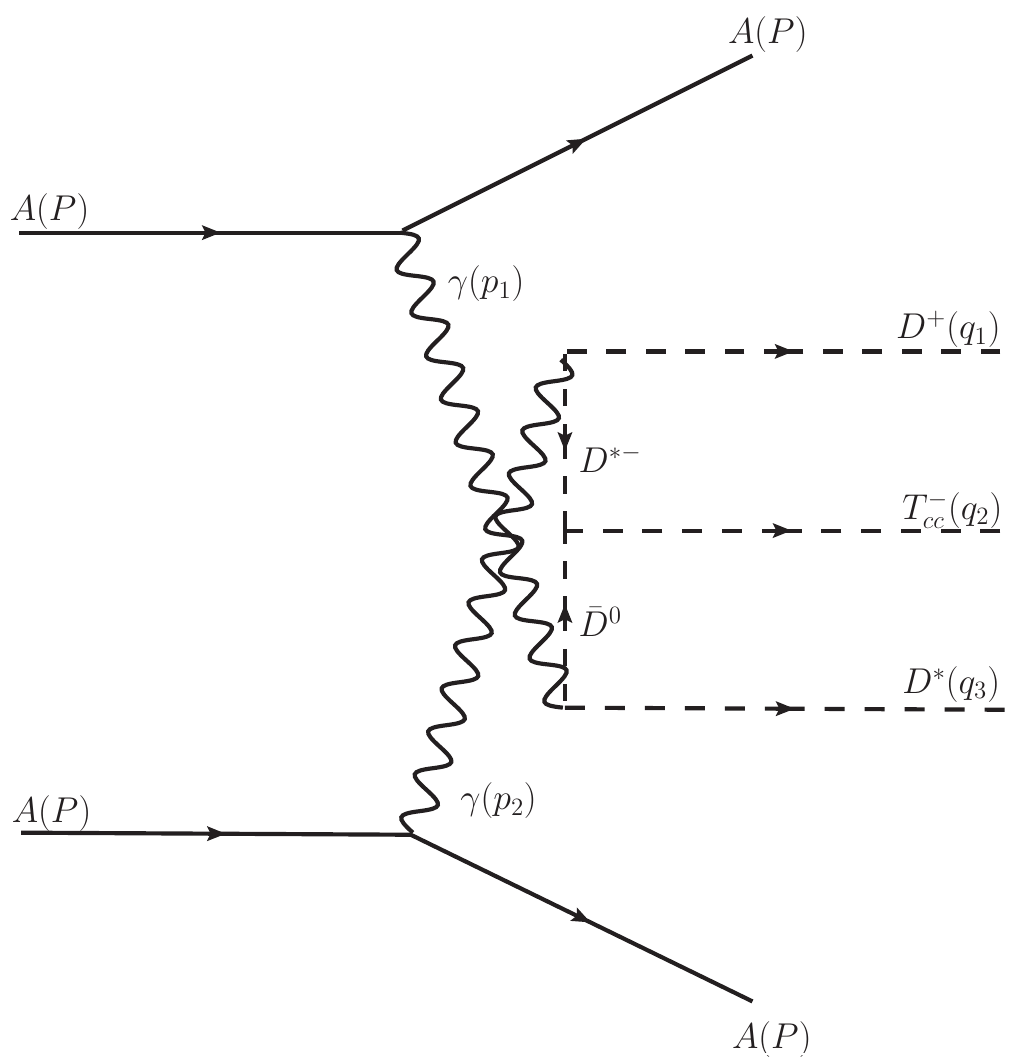}}
	\caption{Two-photon production process for $T^-_{cc}$ in AA, pA or pp UPCs.}
	\label{fig:double_gama}
\end{figure}
The Feynman diagram for the production of $T^-_{cc}$ in two-photon process of UPCs is plotted in Fig.~\ref{fig:double_gama}.  The relevant differential cross-section
is expressed as
 \begin{equation}
 \frac{d\sigma_{AB}}{d{\rm W}_{\gamma \gamma}}=\frac{d\mathcal{L}^{AB}_{\gamma\gamma}}{d{\rm W}_{\gamma \gamma }} \sigma_{\gamma \gamma\rightarrow D^+ T^-_{cc} D^*}({\rm W}_{\gamma\gamma}), \label{eq:two_photon_diff_cross_UPCs}
 \end{equation}
where A and B represent the nucleus or proton. $\sigma_{\gamma \gamma\rightarrow D^+ T^-_{cc} D^*}$ denotes the total cross-section of the two-photon $T^-_{cc}$
production process, and ${\rm W}_{\gamma\gamma}$ stands for the center-of-mass energy of the two-photon system. The effective two-photon luminosity, denoted as $\frac{d\mathcal{L}^{AB}_{\gamma\gamma}}{d{\rm W}_{\gamma \gamma }}$, can be obtained from the gamma-UPC package \cite{Shao:2022cly}.  However, the two-photon
$T^-_{cc}$ production cross-section is unknown and will be discussed later.

To compute the two-photon $T^-_{cc}$ production cross-section $\sigma_{\gamma \gamma\rightarrow D^+ T^-_{cc} D^*}$, the effective Lagrangians with the smallest number
of derivatives are given as follows \cite{GUO2015394,RevModPhys.90.015004,PhysRevD.104.116008,Du:2020bqj,PhysRevD.97.094013}
 \begin{equation}
 \begin{split}
 \mathcal{L}_{T_{c c} D^* D}&=g_{T_{c c}D^*D} T_{c c}^{\mu \dagger} D_\mu^* D,\\
 \mathcal{L}_{\gamma D D^*}&=g_{\gamma D D^*} \epsilon_{\mu \nu \alpha \beta}\left(\partial^\mu \mathcal{A}^\nu\right)\left(\partial^\alpha D^{* \beta}\right) D+\text { H.c.},
 	\end{split}
 \label{eq:L_effect}
 \end{equation}
where $T_{cc}^\mu$, $D^*_\mu$, $D$ and $\mathcal{A^{\mu}}$ represent $T^{-}_{cc}$ meson, $D^*$ meson, $D$ meson, and the photon, respectively.

Then, as shown in Fig.$\,$\ref{fig:subfig:double_gama1} and \ref{fig:subfig:double_gama2} for the two-photon $T^-_{cc}$ production, the invariant amplitude of $\gamma \gamma\rightarrow D^+ T^-_{cc} D^*$ is written as
\begin{equation}
	-i\mathcal{M}=\epsilon^*_\theta (q_3)\epsilon^*_\mu (q_2)(\mathcal{W}^{\theta \mu \nu \alpha }_{(a)}+ \mathcal{W}^{\theta \mu \nu \alpha }_{(b)} )\epsilon_\nu(p1)\epsilon_\alpha(p2).
	\label{eq:two-gama amp}
\end{equation}
In the amplitude,  $\mathcal{W}^{\theta \mu \nu \alpha }_{(a)}$ and  $\mathcal{W}^{\theta \mu \nu \alpha }_{(b)}$ are similar. They are constructed as
\begin{equation}
	\begin{split}
			\mathcal{W}^{\theta \mu \nu \alpha}_{(a)}=&g_aF_a\epsilon_{\beta \nu \eta \rho }p_1^{\beta}k^{\eta}_1\frac{-g^{\mu \rho}+k^{\mu}_1k^{\rho}_1/m^2_{D^{*-}}}{k^2_1-m_{D^{*-}}^2}\times\\
			&\frac{1}{k^2_2-m^2_{\bar{D}^0}}\epsilon_{e\alpha f \theta}p_2^e q^f_3,
	\end{split}
\label{eq:w}
\end{equation}

\begin{equation}
	\begin{split}
		\mathcal{W}^{\theta \mu \nu \alpha}_{(b)}=&g_aF'_a\epsilon_{\beta \nu \eta \rho }p_2^{\beta}k'^{\eta}_1\frac{-g^{\mu \rho}+k'^{\mu}_1k'^{\rho}_1/m^2_{D^{*-}}}{k'^2_1-m_{D^{*-}}^2}\times\\
		&\frac{1}{k'^2_2-m^2_{\bar{D}^0}}\epsilon_{e\alpha f \theta}p_1^e q^f_3,
	\end{split}
	\label{eq:w}
\end{equation}
where $g_a=g_{T_{c c}D^{*-}\bar{D}^0} g_{\gamma D^+ D^{*-}} g_{\gamma \bar{D}^0 D^{*0}}$ and ${\cal{F}}_a={\cal{F}}_{D^{*-}}{\cal{F}}_{\bar{D}^0}$.


The differential cross-section for two-photon $T^-_{cc}$ production can be described as follows
\begin{equation}
	\begin{split}
			d \sigma_{\gamma\gamma}=&\frac{1}{(2 \pi)^{5}} \frac{1}{4 \sqrt{\left.(p_{1} \cdot p_{2}\right)^{2}}} \overline{|\mathcal{M}|}^{2}\times\\ &\delta^{4}\left(p_{1}+p_{2}-q_{1}-q_{2}-q_{3}\right) \frac{d^{3} \vec{q}_{1}}{2 q_{1}^{0}} \frac{d^{3} \vec{q}_{2}}{2 q_{2}^{0}} \frac{d^{3} \vec{q}_{3}}{2 q_{3}^{0}}.
	\end{split}
\label{eq:two_photon_diff_cross}
\end{equation}
where $\overline{|\mathcal{M}|}$ represents the final-state spin summation and the initial-state spin averaging of the scattering amplitude.
Finally, the total cross section, obtained by integrating the differential cross-section with the 3BodyXSections package~\cite{timmurphy.org},
will be presented as a function of the center-of-mass energy $\rm W$.

\section{Numerical results}
\label{sec:numerical redult}
In this study, we estimate the production cross-section of $T^{-}_{cc}$ through one-photon and two-photon processes in ultra-peripheral collisions (UPCs).
To achieve these results, we need to initially evaluate the cross-sections for the $\gamma \gamma \rightarrow D^+ T^{-}_{cc} D^*$ and $\gamma p \rightarrow D^{+} T^{-}_{cc} \Lambda^{+}_{c}$ reactions, as indicated in Eqs.~(\ref{eq:single__gama_diff_sigma}) and (\ref{eq:two_photon_diff_cross_UPCs}). These cross-sections can be
computed by integrating the differential cross-section based on the 3BodyXSections package~\cite{timmurphy.org}.  The equations expressing the differential
cross-section for these two reactions can be found in Eqs.~(\ref{eq7}) and (\ref{eq:two_photon_diff_cross}).

\begin{figure}[htbp]
	\centering
	\includegraphics[scale=0.9]{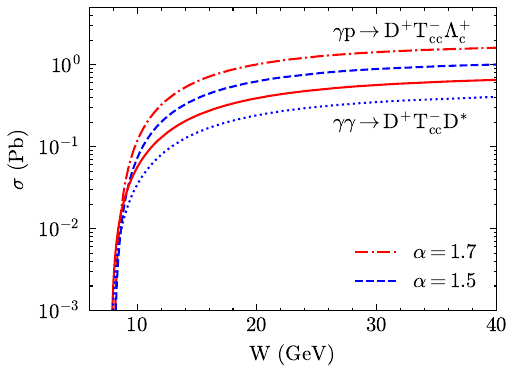}
	\caption{Total cross sections for the $\gamma p \rightarrow D^+ T^-_{cc}\Lambda^{+}_{c}$ (Dotted line and Dashed line) and for the $\gamma \gamma\rightarrow D^+ T^-_{cc} D^*$ (Solid line and Dot line) as a function of $\rm W$ while $\alpha=1.7$ (red line) or $\alpha=1.5$ (blue line). }
	\label{fig:sigma_gaga}
\end{figure}

The cross-sections of the $\gamma p \rightarrow D^{+} T^{-}_{cc} \Lambda^{+}_{c}$ and the $\gamma \gamma \rightarrow D^+ T^{-}_{cc} D^{*}$ reactions are 
depicted against $\rm W$ for $\alpha=1.5$ and $\alpha=1.7$ in Fig.~\ref{fig:sigma_gaga} respectively.  We can find 
that when the energy approaches the $D^{+} T^{-}_{cc} D^{*}$ threshold,  the total cross-section increases sharply.  At higher energies, the cross section 
increases continuously but relatively slowly compared with that near threshold.  Our numerical results also show that the total cross section for $\alpha=1.7$ 
is larger that of the $\alpha=1.5$, but the disparity is not significant.  To provide an example, let's examine the cross-section at an energy of around 
$\rm W=40$ GeV. In this instance, the obtained cross-section ranges from 1.0 Pb to 1.62 Pb for $\gamma p \rightarrow D^{+} T^{-}_{cc} \Lambda^{+}_{c}$ and 
from 0.405 pb to 0.655 pb for $\gamma \gamma \rightarrow D^+ T^{-}_{cc} D^{*}$, respectively, when altering the value of $\alpha$ from 1.5 to 1.7.  Therefore, 
in the following calculations, we only provide the results with $\alpha=1.5$.  By comparing the cross-sections depicted in Fig.~\ref{fig:sigma_gaga}, we find that the total cross-section for $T^{-}_{cc}$ production in the $\gamma p \rightarrow D^+ T^-_{cc}\Lambda^{+}_{c}$ reaction is bigger than that of the $T^{-}_{cc}$ production in the $\gamma \gamma \rightarrow D^+ T^{-}_{cc} D^{*}$ reaction.  

\begin{figure*}[htbp]
	\centering
	\subfigure[\label{fig:subfig:single_gama_Pbp}]{
		\includegraphics[scale=0.65]{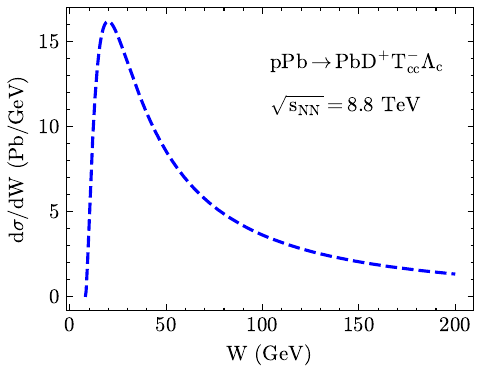}}
	\hspace{0.0098in}
	\subfigure[\label{fig:subfig:single_gama_Aup}]{
		\includegraphics[scale=0.65]{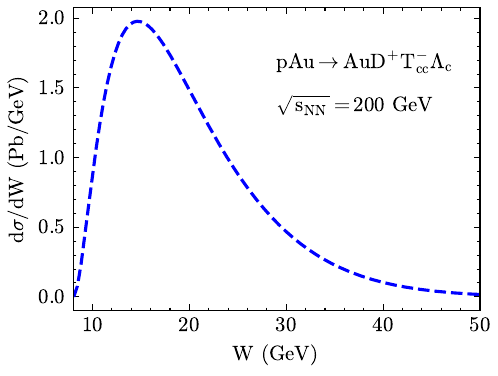}}
	\hspace{0.0098in}
	\subfigure[\label{fig:subfig:single_gama_pp}]{
		\includegraphics[scale=0.65]{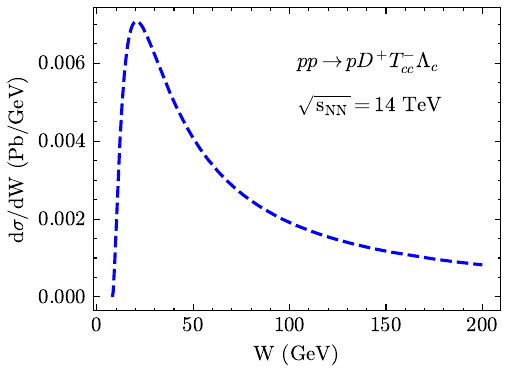}}
	\caption{Differential cross section as a function of $\rm W$ of one-photon $T^{-}_{cc}$ production while subprocess is $\gamma p \rightarrow D^{+}T^{-}_{cc}\Lambda_{c}$ in p-Pb, p-Au, p-p UPCs respectively (from left to right).}
	\label{fig:single_gama_dsigma_dw_Pbp_Aup}
\end{figure*}

\begin{figure*}[htbp]
	\centering
		\subfigure[\label{fig:subfig:two_gama_pPb}]{
		\includegraphics[scale=0.65]{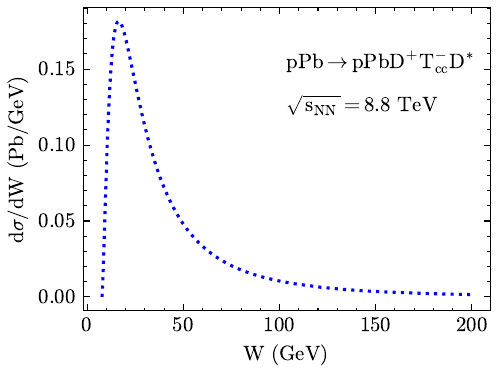}}
	\hspace{0.098in}
	\subfigure[\label{fig:subfig:two_gama_pAu}]{
		\includegraphics[scale=0.65]{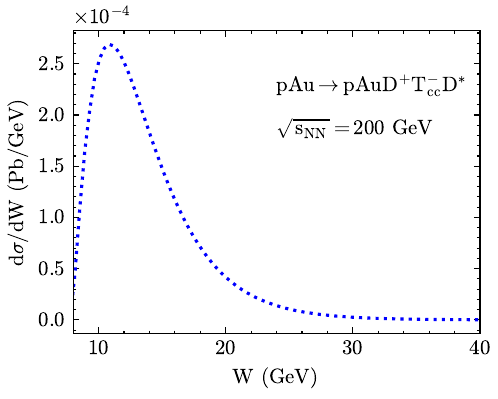}}
		\hspace{0.098in}
	\subfigure[\label{fig:subfig:two_gama_pp}]{
		\includegraphics[scale=0.65]{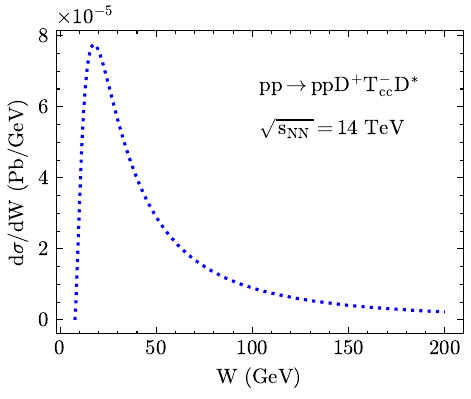}}
			\hspace{0.0098in}
	\subfigure[\label{fig:subfig:two_gama_PbPb}]{
		\includegraphics[scale=0.65]{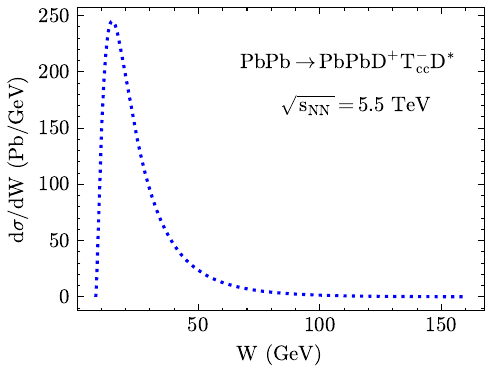}}
	\hspace{0.0098in}
	\subfigure[\label{fig:subfig:two_gama_OO}]{
		\includegraphics[scale=0.65]{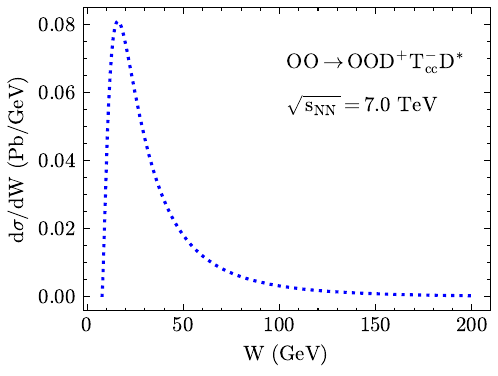}}
	\centering
	\caption{The differential cross section $d\sigma/d\rm W$ of two-photon $T^{-}_{cc}$ production while subprocess is $\gamma \gamma \rightarrow D^{+}T^{-}_{cc}D^*$ in p-Pb, p-Au, p-p, Pb-Pb and O-O UPCs respectively.}
	\label{fig:two_gama_dsigma_dw_AA}
\end{figure*}

\begin{table*}[htbp]\small
	\centering
	\caption{1. Integrated luminosity per typical run $\mathcal{L}_{\rm int}$ and c.m. energy $\sqrt{s_{NN}}$ of nucleus (proton) -nucleus (proton) UPCs from HL-LHC and RHIC \cite{Shao:2022cly,Bruce_2020,dEnterria:2022sut,ParticleDataGroup:2018ovx}. 2. The total cross sections and the events numbers of one-photon and two-photon $T^{-}_{cc}$ production in different kinds of UPCs. }	
	\label{tab:parameter of wave function}       %
		\begin{tabular*}{0.9\textwidth}
			{m{0.15\textwidth}m{0.1\textwidth}m{0.144\textwidth}m{0.125\textwidth}m{0.12\textwidth}m{0.115\textwidth}m{0.09\textwidth}}
			\hline\noalign{\smallskip}
			\centering
			\multirow{2}*{system}&\multirow{2}*{$\sqrt{s_{NN}}$}&\multirow{2}*{$\mathcal{L}_{\rm int}\,\rm (pb^{-1})$}&\multicolumn{2}{c}{$\sigma_{\rm tot} \,(\rm pb)$}&\multicolumn{2}{c}{events}\\
			\cline{4-7}
			\centering
			~& ~& ~&one-photon & two-photon&one-photon&two-photon\\
			
			\noalign{\smallskip}\hline\noalign{\smallskip}
			\centering
			Pb-Pb&$5.5 \,\rm TeV$&$5\times 10^{-3}$&-&5191&-&25.9 \\
			
			\noalign{\smallskip}\hline\noalign{\smallskip}
			\centering
			O-O&$7.0 \,\rm TeV$&12&-&2.5&-&30 \\
			
			\noalign{\smallskip}\hline\noalign{\smallskip}
			\centering
			Pb-p&$8.8 \,\rm TeV$&1&1000&6.0&1000&6.0 \\
			
			\noalign{\smallskip}\hline\noalign{\smallskip}
			\centering
			Au-p&$200 \,\rm GeV$&4.5&30.1&0.002&135.5&$<1$ \\
			
			\noalign{\smallskip}\hline\noalign{\smallskip}
			\centering
			p-p&$14 \,\rm TeV$&$1.5\times 10^{5}$&0.48&0.0035&$7.2\times10^{4}$&525 \\		
			
			\noalign{\smallskip}\hline	
		\end{tabular*}
	\label{tab:enents}
\end{table*}
With above obtained cross-section, we present the differential cross-sections for one-photon $T^{-}_{cc}$ production in ultra-peripheral collisions (UPCs) 
as a function of the $\rm W$ of the photon-nucleus system at $\sqrt{s}=8.8$ TeV for the $\rm p-Pb$ system and $\sqrt{s}=200$ GeV for the $\rm p-Au$ system, as 
illustrated in Fig.~\ref{fig:single_gama_dsigma_dw_Pbp_Aup}.  We can find that the differential cross-sections for the $\rm p-Pb$ system are greater than those 
of the $\rm p-Au$ system.   One possible explain for this is that the photon flux is directly proportional to the charge number of the nucleus. Moreover, our results 
suggest that the differential cross-sections are notably significant in the low-energy range.  As the energy increases, the differential cross-sections decrease 
rapidly.

Next, we calculate the differential cross-sections through two-photon $T^{-}_{cc}$ production in nucleus-nucleus and proton-proton ultra-peripheral collisions (UPCs), 
as illustrated in Fig.~\ref{fig:two_gama_dsigma_dw_AA}. These calculations are performed for collision energies of $\sqrt{s}=5.5\,\rm TeV$ for the $\rm Pb-Pb$ system, 
$\sqrt{s}=7$ TeV for the oxygen-oxygen ($\rm O-O$) system, and $\sqrt{s}=14$ TeV for the $\rm p-p$ system, respectively. Moreover, for the $\rm p-Pb$ system and $\rm p-Au$ system, 
we present the differential cross-sections against the center-of-mass energy $\rm W$, depicted in Fig.~\ref{fig:two_gama_dsigma_dw_AA}, for collision energies of 
$\sqrt{s}=8.8$ TeV and $\sqrt{s}=200$ GeV, respectively.  It can be observed that the primary contribution of the two-photon process takes place within a distinct low 
center-of-mass energy range, similar to that of one-photon $T^{-}_{cc}$ production.    Notably, $T^{-}_{cc}$ production in $\rm Pb-Pb$ UPCs is the largest due to the higher luminosity of the photon flux originating from Pb.

Finally, we list the predicted event numbers for one-photon and two-photon $T^{-}_{cc}$ production in the $\rm Pb-Pb$ system, $\rm O-O$ system, $\rm Pb-p$ system, $\rm Au-p$ system, and $\rm p-p$ system, respectively, in Table.~\ref{tab:enents}. We can find that the total cross-section $\sigma_{\rm tot}$ for the one-photon $\gamma p\rightarrow D^+ T^{-}_{cc}\Lambda^+_{c}$ process is smaller in the $\rm p-p$ system compared to the $\rm Pb-p$ and $\rm Au-p$ systems.  However, due to the higher integrated luminosity per typical run $\mathcal{L}_{\rm int}$ in the $\rm p-p$ system, the event number is larger than that of the other processes.  Conversely, it is noticeable that in the case of the two-photon process, the total cross-sections are smaller compared to those of the one-photon process in the $\rm Pb-p$, $\rm Au-p$, and $\rm p-p$ systems.  Despite the total cross section $\sigma_{\rm tot}$ reaching approximately 5 nb 
for the two-photon $T^{-}_{cc}$ process in the $\rm Pb-Pb$ system, owing to the very high two-photon number densities, the restricted luminosity results in a relatively small number 
of events. As a result, identifying $T^{-}_{cc}$ through the one-photon process is more feasible.


\section{CONCLUSION}
\label{sec:summary}
In this work, theoretical frameworks of one-photon and two-photon $T^{-}_{cc}$ production are introduced. Then the differential distribution and the total cross section of one-photon and two-photon $T^{-}_{cc}$ production in UPCs are calculated. At last, the events of $T^{-}_{cc}$ production are estimated in different collision systems.

Based on the work of \cite{PhysRevD.104.116008} for $\gamma p\rightarrow D^{+}T^{-}_{cc}\Lambda_{c}$, the differential cross sections as a function of the center-of-mass energy of photon and proton ($\rm W$) in Pb-p, Au-p and p-p collisions are calculated. And Referring to $\gamma p\rightarrow D^{+}T^{-}_{cc}\Lambda_{c}$ process, the t-channel amplitude of $\gamma \gamma \rightarrow D^{+}T^{-}_{cc}D^*$ is presented. Then the differential cross sections for two-photon $T^{-}_{cc}$ UPCs in different collision systems are shown. At last, the events of one-photon and two-photon $T^{-}_{cc}$ production are estimated respectively. Due to high-luminosity photon flux in Pb-p and Pb-Pb systems,  the total cross section of one-photon UPC process for $T^{-}_{cc}$ production is approximately 1$\,$nb and the total cross section of two-photon UPC process for $T^{-}_{cc}$ production is approximately 5$\,$nb respectively. But because of the limited integrated luminosity in Pb-p and Pb-Pb systems, the number of events is lower. In p-p system, despite the lower production cross-section of $T^{-}_{cc}$ , the higher number of events is generated due to that the integrated luminosity is about $1.5\times 10^{5}\,\rm Pb^{-1}$. In conclusion,  It is more possible to identify $T^{-}_{cc}$ in p-p UPCs in the future HL-LHC.
\begin{acknowledgments}
This work is supported by
the Strategic Priority Research Program of Chinese Academy of Sciences under the Grant NO. XDB34030301.
\end{acknowledgments}

\bibliographystyle{apsrev4-1}
\bibliography{refs_INSPIRE}

\end{document}